\documentstyle[12pt]{article}
\date{}
\def\be{\begin{equation}}
\def\ee{\end{equation}}
\def\bea{\begin{eqnarray}}
\def\eea{\end{eqnarray}}
\def\s{\sigma}
\def\al{\alpha}

\def\de{\delta}
\def\om{\omega}
\def\pr{\prime}
\def\th{\theta}

\title{Excited states of rotating\\
relativistic string with massive ends}
\author{G.\,S. Sharov\\
{\small Tver state university}\\
{\small Tver, 170002, Sadovyj per. 35, Mathem. dep-t.}}
\begin{document}
\maketitle
\begin{abstract}
For the relativistic string with massive ends arbitrary small disturbances
of the uniform rotation of the rectilinear string are investigated.
There are two classes of these oscillations with different spectra of
frequencies. They are stationary waves oscillating in the rotational plane
and in the orthogonal direction.
These states are presented in the form of Fourier series that helps
to quantize string excitations in the linear vicinity of the classical
rotation. These string oscillations may be applied to describing daughter
Regge trajectories in the framework of the considered model.
\end{abstract}

\bigskip
\noindent{\bf Introduction}
\medskip

The relativistic string with massive ends \cite{Ch,BN} is considered
as the string model of the meson $q$-$\overline q$ or the quark-diquark model
$q$-$qq$ of the baryon. This model includes two massive
points connected by the string simulating strong interaction between quarks
and the QCD confinement mechanism.

Many authors \cite{Ko,Solov,4B,InSh} used this model to describe orbitally
excited states of mesons and baryons lying on (quasi)linear Regge
trajectories. On the classical level these states correspond to
uniform rotations of the system and such rotations of the string with massive
ends generate the quasilinear Regge trajectories in the natural way.

But for describing
the higher radial excitations and other hadron excited states we are to use
more complicated string motions than rotational ones.
There were numerous attempts to expand the application area of this string
model and to specify required states among the whole variety of string motions.
From this point of view small disturbances of the string rotational motions
are very interesting.

In particular, some authors tried to obtain these disturbances of rotating
string in the framework of fixed rectilinear shape of the string
\cite{Ida,Olss89}.
But it was shown in Ref.~\cite{Nestrad} that the required disturbed motions
must have curved shape of the rotating string with massive ends.

Taking this point into account the authors of Refs.~\cite{AllenOV}
searched small disturbances of the uniform rotation for the
string with the fixed end in the form
\be
X^0=t,\quad
X^1+iX^2=\s R(t)\cdot\exp{\left[i\big(\Omega t+\phi(t)+f(\s)\big)\right]},
\quad\s\in[0,1].
\label{AOV}\ee
Here $X^\mu(\tau,\s)$ is the world surface of the relativistic string in  Minkowski space $R^{1,2}$ where the evolution parameter $\tau$ is equal to time:
$\tau\equiv t$.

In the case $R(t)=R_0={}$const, $\phi(t)=0$, $f(\s)=0$
the expression (\ref{AOV}) described the rotational motion
(uniform rotation of the rectilinear string at the angular frequency $\Omega$)
and it is the exact solution of the dynamical equations of the relativistic
string \cite{BN}
\be
\frac\partial{\partial\tau}\frac{\partial\sqrt{-g}}{\partial\dot X^\mu}
+\frac\partial{\partial\s}\frac{\partial\sqrt{-g}}{\partial X^{\pr\mu}}=0
\label{eqgen}\ee
satisfying the boundary conditions (or equations of motion for endpoints)
\be
\frac d{d\tau}\frac{\dot x_{i\mu}}{\sqrt{\dot x_i^2}}-\frac{(-1)^i\gamma}{m_i}
\bigg[\frac{\partial\sqrt{-g}}{\partial X^{\pr\mu}}-\s'_i(\tau)\,
\frac{\partial\sqrt{-g}}{\partial\dot X^\mu}\bigg]\bigg|_{\s=\s_i}\!\!\!=0,\quad
i=1,2.
\label{qqgen}\ee
Here $-g=(\dot X,X')^2-\dot X^2X'{}^2$,
$\dot X^\mu=\partial_{\tau}X^\mu$, $X^{\pr\mu}=\partial_\sigma X ^\mu$,
$(a,b)=a_\mu b^\mu$ is the (pseudo)scalar product in Minkowski space $R^{1,3}$
with signature $(+,-,-,\dots)$, $\gamma$ is the string tension,
the speed of light $c=1$,
$\dot x_i^\mu(\tau)=\frac d{d\tau}X^\mu(\tau,\s_i(\tau))$; $m_i$ are the
masses of the endpoints, $\s_i(\tau)$
are inner coordinates of their world lines. These massive points
simulate the quark-antiquark pair for the meson or the quark and diquark
for the baryon model $q$-$qq$.

Equation (\ref{eqgen}) and conditions (\ref{qqgen}) result from the action \cite{BN}
\be
 S=-\gamma\!\int\sqrt{-g}\;d\tau\,d\s+\sum_{i=1}^2m_i\int
\sqrt{\dot x_i^2(\tau)}\;d\tau,
\label{S}\ee
and also from the action in Refs.~\cite{AllenOV}, where in the first summand
the Lagrangian
$L=-\frac12\gamma h^{\al\beta} X^\mu_{,\al}X_{\mu,\beta}\sqrt{-h}$
in the Polyakov's form \cite{Pol81} was used.

The authors of Refs. \cite{AllenOV} substituted the ansatz (\ref{AOV})
into Eq.~(\ref{eqgen}) and obtained its particular solution,
assuming that the disturbances $R(t)-R_0$, $\phi(t)$, $f(\s)$ are small.
But they ignored the boundary condition (\ref{qqgen}) and the essential
dependence of the obtained function $f(\s)$ on $t$.

It was shown in Ref.~\cite{stabPRD} that the mentioned assumptions were
incorrect and the adequate general solution of this problem was suggested.
Arbitrary small disturbances of rotational motions for the
relativistic string with one massive and one fixed ends were described.

In this paper the mentioned approach applied to the string with two massive
ends. The orthonormality conditions on the string world surface and the
methods suggested in Refs.~\cite{An,PeSh} were used.
In Sect.~1  the nonlinear dynamics of this system with Eqs.~(\ref{eqgen}),
(\ref{qqgen}) was reduced to the system of ordinary differential equations
with shifted arguments.
In Sect.~2 small disturbances of the rotational motion for the string with
massive ends (the so called quasirotational states) are described.
In Sect.~3 the approach to quantization of these states is suggested
and possibilities for describing the higher radial excitations of hadrons
are discussed.

\bigskip
\noindent{\bf 1. Dynamics of the string with massive ends}
\medskip

Unlike the authors \cite{AllenOV} we use more convenient approach and
consider the string dynamics under the orthonormality conditions
on the world surface
\be
(\dot X\pm X')^2=0.
\label{ort}\ee
These gauge conditions and also the requirements
\be
\s_1=0,\qquad\s_2=\pi.
\label{s0pi}\ee
always may be stated without loss of generality \cite{BN,stabPRD}
if we use the invariance of action (\ref{S}) with respect to nondegenerate
reparametrizations on the world surface.

The equations of motion (\ref{eqgen}) and the boundary conditions at the ends
(\ref{qqgen}) under the conditions (\ref{ort}) and (\ref{s0pi})
take the following simplest form
\bea
&\ddot X^\mu-X''{}^\mu=0,&\label{eq}\\
&\displaystyle
m_i\frac d{d\tau}U^\mu_i(\tau)+(-1)^i\gamma
X^{\pr\mu}(\tau,\s_i)=0,\quad i=1,2.&
\label{qq}\eea
Here
\be
U^\mu_i(\tau)=\frac{\dot x_i^\mu(\tau)}{\sqrt{\dot x_i^2(\tau)}}=
\frac{\dot X^\mu(\tau,\s_i)}{\sqrt{\dot X^2(\tau,\s_i)}}
\label{Ui}\ee
is the unit $R^{1,D-1}$ velocity vector of the $i$-th string end.

The equations (\ref{eq}) are linear
but the boundary conditions (\ref{qq}) for the massive points
remain essentially nonlinear.

If we substitute the general solution of equations (\ref{eq})
\be
X^\mu(\tau,\s)=\frac1{2}\Big[\Psi^\mu_+(\tau+\s)+\Psi^\mu_-(\tau-\s)\Big]
\label{gen}\ee
into the boundary conditions (\ref{qq}),
the problem is reduced to the system of ordinary differential equations
with shifted arguments.
The unknown function may be $\Psi^\mu_+(\tau)$, $\Psi^\mu_-(\tau)$, or
unit velocity vectors (\ref{qq}) of the endpoints $U^\mu_1(\tau)$ or $U^\mu_2(\tau)$
--- this is equivalent due to the relations \cite{An,PeSh}
\be
\Psi^{\pr\mu}_\pm(\tau\pm\s_i)=m_i\gamma^{-1}\Big[
\sqrt{-U_i^{\pr2}(\tau)}\,U_i^\mu(\tau)\mp(-1)^i U_i^{\pr\mu}(\tau)\Big].
\label{psdet}\ee
Remind that $\s_1=0$, $\s_2=\pi$.

Taking relations (\ref{ort}) and (\ref{gen}) into account we transform
the boundary conditions (\ref{qq}) into the systems \cite{PeSh}
$$
\frac{dU_i^\mu}{d\tau}=\mp(-1)^i\frac\gamma{m_1}\big[\de^\mu_\nu-
U_i^\mu(\tau)\,U_{i\nu}(\tau)\big]\,\Psi^{\pr\nu}_\pm(\tau\pm\s_i),
$$
where $\de^\mu_\nu=\left\{\begin{array}{cl} 1, & \mu=\nu\\ 0, & \mu\ne\nu.
\end{array}\right.$

Substituting Eqs.~(\ref{psdet}) into these relations we obtain for the case
$0<m_i<\infty$ the system
\be
\begin{array}{c}
U^{\pr\mu}_1(\tau)=m_2m_1^{-1}\big[\de^\mu_\nu-
U_1^\mu(\tau)\,U_{1\nu}(\tau)\big]\big[\sqrt{-U_2^{\pr2}(\tau-\pi)}\,
U_2^\nu(\tau-\pi)-U_2^{\pr\nu}(\tau-\pi)\big],\\
U^{\pr\mu}_2(\tau)=m_1m_2^{-1}\big[\de^\mu_\nu-
U_2^\mu(\tau)\,U_{2\nu}(\tau)\big] \big[\sqrt{-U_1^{\pr2}(\tau-\pi)}\,
U_1^\nu(\tau-\pi)-U_1^{\pr\nu}(\tau-\pi)\big],
\rule[3.5mm]{0mm}{1mm}\end{array}
\label{sysU}\ee
that plays the role of the above mentioned system of ordinary differential equations
with shifted arguments with respect to unknown  vector functions $U^\mu_i(\tau)$.
Note that the initial data (keeping all information about the string's motion)
may be given here as the function $U^\mu_1(\tau)$ or $U^\mu_2(\tau)$
in the segment $I=[\tau_0,\tau_0+2\pi]$ with the parameters $m_i/\gamma$,
$U^\mu_i(\tau_0+\pi)$. Integration of the system (\ref{sysU}) with this initial
data yields the values $U^\mu_1(\tau)$ and $U^\mu_2(\tau)$ for all $\tau$
and then we can obtain the world surface $ X^\mu(\tau,\s)$ with the help of
Eqs.~(\ref{psdet}) and (\ref{gen}).

\bigskip
\noindent{\bf 2. Quasirotational states }
\medskip

For the relativistic string with massive ends the classical rotational motion
(planar uniform rotation of the
rectilinear string segment) is well known \cite{Ch,Ko} and may be represented
in the form \cite{4B,stabPRD}
\be
X^\mu(\tau,\s)=X^\mu_{rot}(\tau,\s)=\Omega^{-1}\big[\th\tau e_0^\mu+
\cos(\th\s+\phi_1)\cdot e^\mu(\tau)\big].
\label{rot}\ee
Here $\Omega$ is the angular velocity,
$e_0^\mu$ is the unit time-like velocity vector of c.m.,
\be
e^\mu(\tau)=e_1^\mu\cos\th\tau+e_2^\mu\sin\th\tau
\label{evec}\ee
is the unit ($e^2=-1$)
space-like rotating vector directed along the string,
$\s\in[0,\pi]$. The parameter $\th$ (dimensionless frequency) is connected
with the constant speeds $v_1$, $v_2$ of the ends through the relations
\be
v_1=\cos\phi_1,\quad v_2=-\cos(\pi\th+\phi_1),\quad
\frac{m_i\Omega}\gamma=\frac{1-v_i^2}{v_i}.
\label{vom}\ee

Expression (\ref{rot}) under restrictions (\ref{vom}) is the exact solution of the classic
equations of motion (\ref{eq}) and satisfies the orthonormality (\ref{ort}) and boundary conditions (\ref{qq}).
The classic expressions for the energy $E$ and the angular momentum $J$ (its projection
onto $Oz$ or $e_3^\mu$-direction) of the states (\ref{rot}) are \cite{Ch,Ko,4B}
\be
E_{rot}=\sum_{i=1}^2\bigg[\frac{\gamma\arcsin v_i}\Omega+
\frac{m_i}{\sqrt{1-v_i^2}}\bigg],\qquad
J_{rot}=\frac1{2\Omega}\bigg\{\sum_{i=1}^2\bigg[\frac{\gamma\arcsin v_i}\Omega+
\frac{m_iv_i^2}{\sqrt{1-v_i^2}}\bigg]\bigg\},
\label{EJ}\ee
The implicit expression (\ref{EJ}) with different form of taking into account quark spins
and the spin-orbit correction \cite{Ko,4B} describes quasilinear Regge trajectories
$J=J(E^2)$ with the ultrarelativistic behavior
$J\simeq\alpha'E^2-\alpha_1 E^{1/2}$, $E\to\infty$,
where the slope has the Nambu value $\alpha'=(2\pi\gamma)^{-1}$.
So the rotational motions  of the string models $q$-$\overline q$ and $q$-$qq$
are widely used for modeling the orbitally excited hadron
states on the leading (parent) Regge trajectories \cite{Ko,Solov,4B,InSh}.

But other excited hadron states, for example, the radial excitations on the
daughter Regge trajectories \cite{InSh} are so far beyond the field
of application for the string model (\ref{S}).
For this purpose small disturbances of the rotational motion (\ref{rot})
were searched in Refs.~\cite{Ida,Olss89,AllenOV}.
But these attempts were not fruitful (see Introduction).

These small disturbances of the rotation (\ref{rot}) (we use below the term
``quasirotational states" for these disturbances)
are interesting due to the following reasons:
(a) we are to search the motions describing the radially excited hadron states,
in other words, we are to describe the daughter Regge trajectories;
(b) the quasirotational motions are necessary for solving the problem
of stability of rotational states (\ref{rot});
(c) the quasirotational states are the basis for quantization of these
nonlinear problems in the linear vicinity of the solutions (\ref{rot})
(if they are stable).

In Ref.~\cite{stabPRD} the approach to describe quasirotational states
was suggested for the relativistic string with massive and fixed ends
($0<m_1<\infty$,
$m_2\to\infty$). Here we generalize it for the string with two massive ends.

For the rotational motion (\ref{rot}) the unit velocity vectors $U_i^\mu$
(\ref{Ui}) are
\be
U_i^\mu(\tau)=U_{i(rot)}^\mu(\tau)=\Gamma_i\big[e_0^\mu-(-1)^iv_i
\acute e^\mu(\tau)\big],\qquad\quad\Gamma_i=(1-v_i^2)^{-1/2},
\label{Urot}\ee
where $\acute e^\mu(\tau)=-e_1^\mu\sin\th\tau+e_2^\mu\cos\th\tau=
\th^{-1}\frac d{d\tau}e^\mu(\tau)$ is the unit rotating vector, connected
with the vector $e^\mu$ (\ref{evec}). Expressions (\ref{Urot}) are solutions
of the system (\ref{sysU}) if the parameters $v_i$, $m_i$, $\theta$ are
related by Eqs.~(\ref{vom}).

To study the small disturbances of the rotational motion (\ref{rot})
we consider arbitrary small disturbances of this motion or of the vectors
(\ref{Urot}) in the form
\be
U^\mu_i(\tau)=U^\mu_{i(rot)}(\tau)+u_i^\mu(\tau),\qquad |u_i^\mu|\ll1.
\label{U+u}\ee
For the exhaustive description of this quasirotational state the disturbance
$u_i^\mu(\tau)$ may be given in the initial segment $I=[\tau_0,\tau_0+2\pi]$.
It is small so we neglect in the linear approximation the second order terms.
The equality $U_i^2(\tau)=1$ for both vectors $U_i^\mu$ and $U_{i(rot)}^\mu$
leads in the linear approximation to the condition
\be
U_{i(rot)}^\mu(\tau)\,u_{i\mu}(\tau)=0.
\label{Uu}\ee

When we substitute the expressions (\ref{U+u}) into the system (\ref{sysU})
and omit the second order terms we obtain the linearized system of equations
describing the evolution of small arbitrary disturbances $u_i^\mu$.
Considering projections of these two vector equations onto the basic vectors
$e_0$, $e$, $\acute e$, $e_3$, we reduce it to the following system of

equations with respect to projections of $u_i^\mu$:
\be
\!\!\!\!\!\!\begin{array}{c}
u'_{10}(\tau)+Q_1u_{10}(\tau)-\Gamma_1Q_1u_{1e}(\tau)=
M_0\big[u'_{20}-Q_2u_{20}+\Gamma_2Q_2u_{2e}\big],\\
\!\!u'_{1e}(\tau)+Q_1u_{1e}(\tau)+\theta v_1^{-1}u_{10}(\tau)=
M_1^{-1}\big[-u'_{2e}-Q_1u_{2e}+N_2^*u'_{20}+N_2u_{20}\big],
\rule[3.3mm]{0mm}{1mm}\!\!\\
u'_{20}+Q_2u_{20}+\Gamma_2Q_2u_{2e}=
M_0^{-1}\big[u'_{10}(-)-Q_1u_{10}(-)-\Gamma_1Q_1u_{1e}(-)\big],
\rule[3.3mm]{0mm}{1mm}\\
\!\!\!\!u'_{2e}+Q_2u_{2e}-\theta v_2^{-1}u_{20}=
M_1\big[-u'_{1e}(-)-Q_2u_{1e}(-)+N_1^*u'_{10}(-)+N_1u_{10}(-)\big],
\rule[3.3mm]{0mm}{1mm}\!\!\\
u'_{1z}(\tau)+Q_1u_{1z}(\tau)=(m_2/m_1)\big[-u'_{2z}+Q_2u_{2z}\big],
\rule[3.3mm]{0mm}{1mm}\\
u'_{2z}+Q_2u_{2z}=(m_1/m_2)\big[-u'_{1z}(-)+Q_1u_{1z}(-)\big].
\rule[3.3mm]{0mm}{1mm}\end{array}\!\!\!
\label{sysu}\ee
Here $Q_i=\Gamma_i\theta v_i={}$const, $(-)\equiv(\tau-2\pi)$, the functions
\be
u_{i0}(\tau)=(e_0,u_i)=e_0^\mu u_{i\mu},\qquad u_{ie}(\tau)=(e,u_i),\qquad
u_{iz}(\tau)=(e_3,u_i)
\label{u0ez}\ee
are the projections of the vectors $u_i^\mu(\tau)$ onto the mentioned basis.
The projections of $u_i^\mu$ onto $\acute e^\mu$ may be expressed through
$u_{i0}$: $(\acute e,u_i)=(-1)^iv_i^{-1}u_{i0}$ due to the equality
(\ref{Uu}). Arguments $(\tau-\pi)$ of the functions $u_{20}$, $u_{2e}$,
$u_{2z}$ are omitted.
The constants in Eqs.~(\ref{sysu}) are
$$\begin{array}{c}
M_0=m_2Q_1/(m_1Q_2),\qquad M_1=m_1\Gamma_1/(m_2\Gamma_2),\\
N_i^*=-(-1)^i(1+Q_{3-i}/Q_i)/\Gamma_i,\qquad
N_i=(-1)^i(Q_{3-i}+Q_i\kappa_i)/\Gamma_i,\qquad\kappa_i=1+v_i^{-2}.
\rule[3.3mm]{0mm}{1mm}\end{array}$$

We shall search solutions of the linearized system (\ref{sysu}) in the form
\be
u_i^\mu=A_i^\mu e^{-i\om\tau}.
\label{uharm}\ee
For the last two equations (\ref{sysu}) (they form the closed subsystem)
solutions in the form (\ref{uharm}) exist only if the dimensionless
frequency $\om$ satisfies the transcendental equation
\be
\frac{\om^2-Q_1Q_2}{(Q_1+Q_2)\,\omega}=\cot\pi\omega.
\label{zfreq}\ee
Equation (\ref{zfreq}) has the countable set of real roots
$\om_n$, $n-1<\om_n<n$, the minimal positive root $\om_1$ is equal to the parameter
$\theta$ in Eq.~(\ref{rot}).
These pure harmonic $z$-disturbances corresponding to various $\om_n$
result in the following correction to the motion (\ref{rot}) [due to
Eqs.~(\ref{psdet}), (\ref{gen}) there is only $z$ or $e_3^\mu$\
component of the correction]:
\be
X^\mu(\tau,\s)=X_{rot}^\mu(\tau,\s)+e_3^\mu
A\cos(\om_n\s+\phi_n)\cdot\cos(\om_n\tau+\varphi_0),
\label{zwave}\ee
Here the amplitude $A$ is to be small in comparison with $\Omega^{-1}$.

Expression (\ref{zwave}) describes oscillating string in the form of
orthogonal (with respect to the rotational plane) stationary waves with
$n$ nodes in the interval $0<\s<\pi$. Note that the string ends are not in
nodes, they oscillate along $z$-axis at the frequency
$\Omega_n=\Omega\om_n/\theta$. The shape
$F=A\cos(\om_n\s+\phi_n)$ of the
$z$-oscillation (\ref{zwave}) is not pure sinusoidal with
respect to the distance $s=\Omega^{-1}\cos(\theta\s+\phi_1)$
from the center to a point ``$\s$":
If $n=1$ this dependence is linear. In this trivial case the motion is
pure rotational (\ref{rot}) with a small tilt of the rotational plane.
But the motions (\ref{zwave}) with excited higher harmonics $n=2,3,\dots$
are non-trivial.

The transcendental equation (\ref{zfreq}) was studied in Ref.~\cite{PeSh}
where we proved that its roots $\om_n$ (with $\om_0=0$) and the functions
in Eqs.~(\ref{zwave})
\be
w_n(\s)=\cos(\om_n\s+\phi_n),\qquad n=0,1,2,\dots,\quad(\om_0=0)
\label{wn}\ee
are correspondingly the eigen-values and eigen-functions
of the boundary-value problem
\be
w''(\s)+\om^2w=0,\;\quad \om^2w(0)+Q_1w'(0)=0,\quad\om^2w(\pi)-
Q_2w'(\pi)=0.
\label{wbvp}\ee
The eigen-values (\ref{wn}) are mutually orthogonal with respect to the scalar
product
\be
\langle f,g\rangle=\int\limits_0^\pi f(\s)\,g(\s)\,d\s+
\frac{f(0)\,g(0)}{Q_1}+\frac{f(\pi)\,g(\pi)}{Q_2}.
\label{scal}\ee

It was proved in Ref.~\cite{PeSh} that the functions $w_n(\s)$,
$n=0,1,2,\dots$ form the complete system in the class $C([0,\pi])$, and
the system $\exp(-i\om_n\tau)$, $n\in Z$ (with $\om_{-n}=-\om_n$) is complete
in the class of function $C(I)$, where $I=[\tau_0,\tau_0+2\pi]$.
So any continuous function $f(\tau)$ given in the segment $I$ may
be expanded in the Fouirer series
\be
f(\tau)=\sum_{n=-\infty}^{+\infty}f_n\exp(-i\om_n\tau),\qquad
\tau\in I=[\tau_0,\tau_0+2\pi].
\label{uFour}\ee

As was mentioned above, all information about any motion of this system
is contained in the function $U_i^\mu(\tau)$ given in the segment $I$.
If we expand any small disturbance $u_{1z}(\tau)$ or $u_{2z}(\tau)$ in the
segment $I$ into the Fourier series (\ref{uFour}), this series will describe
the evolution of the given disturbance for all $\tau\in R$, because
any term in (\ref{uFour}) satisfies the evolution equations (\ref{sysu}).
So any small disturbance of the rotational motion (\ref{rot}) in
$e_3^\mu$ direction may be expanded into the Fourier series with the
oscillatory modes (\ref{zwave}).

This also concerns the quasirotational motions in the rotational plane
$e_1,e_2$. They are determined by the first 4 equations (\ref{sysu}).
If we substitute $u_{i0}=A_i\exp(-i\tilde\om\tau)$,
$u_{ie}=B_i\exp(-i\tilde\om\tau)$ into this system we obtain the following
condition for existance of its non-trivial solutions:
\be
\!\left|\begin{array}{cccc}
Q_1-i\tilde\om &-Q_1 &Q_1+i\tilde\om q_{12} &-
Q_1\\ Q_1/v_1^2  &Q_1-i\tilde\om &-\tilde Q_2-
i\tilde\om q_{12} &Q_1-i\tilde\om \\
Q_1+i\tilde\om &Q_1 &\!(Q_1-i\tilde\om q_{12})\,e^{-2\pi
i\tilde\om}\!\! &Q_1 e^{-2\pi i\tilde\om} \\
\tilde Q_1+i\tilde\om q_{12}^{-1} &Q_2-i\tilde\om &
-Q_2v_2^{-2}e^{-2\pi i\tilde\om}&
(Q_2-i\tilde\om)\, e^{-2\pi i\tilde\om}
\end{array}\right|=0.
\label{deter}\ee
Here $q_{12}=Q_1/Q_2$, $\tilde Q_j=Q_j\kappa_j+Q_{3-j}+i\tilde\om$.

There are eigen-frequencies $\tilde\om=\tilde\om_n=n$, $n\in Z$ satisfying
this equation. But the correspondent functions
$u_i^\mu(\tau)=B_i^\mu\exp(-in\tau)$ after substitution into
Eqs.~(\ref{U+u}), (\ref{psdet}), (\ref{gen}) result in quasirotational
excitations of the motion (\ref{rot}), which may be obtained from
Eq.~(\ref{rot}) through the following reparametrization \cite{PeSh}
$$
\tilde\tau\pm\tilde\s=f(\tau\pm\tilde\s):\quad
f(\xi+2\pi)=f(\xi)+2\pi,\quad f'(\xi)>0,\quad\xi\in R
$$
on the same world surface. Eqs.~(\ref{eq})\,--\,(\ref{ort}) and the conditions
$\s_1=0$, $\s_2=\pi$ are invariant with respect to this reparametrization
so the oscillations with $u_i^\mu(\tau)=B_i^\mu\exp(-in\tau)$ have no
physical manifestations. They may be interpreted as ``longitudinal
oscillations" inside the string or, oscillations of the grid chart on the
world surface.

If we exclude these non-physical roots $\tilde\om_n=n$, the equation
(\ref{deter}) will reduce to following one:
\be
\frac{(\tilde\om^2-q_1)(\tilde\om^2-q_2)-4Q_1Q_2\tilde\om^2}
{2\tilde\om\big[Q_1(\tilde\om^2-q_2)+Q_2(\tilde\om^2-
q_1)\big]}={\mbox{cot}}\,\pi\tilde\om,
\label{pfreq}\ee
where
$$q_i=\theta^2\frac{1+v_i^2}{1-v_i^2}.$$

One can numerate the roots $\tilde\om=\tilde\om_n$ of Eq.~(\ref{pfreq})
in order of increasing so that $\tilde\om_0=0$,
$n-1<\tilde\om_n<n$ for $n\ge1$ (and $\tilde\om_{-n}=-\tilde\om_n$).
The value $\tilde\om=\theta$ is also the root of Eq.~(\ref{pfreq}) but
it corresponds to the trivial quasirotational states, connected with
a small shift of the rotational center with respect to the coordinate origin.

The roots $\tilde\om_n$ of Eq.~(\ref{pfreq}) are eigen-values
of the boundary-value problem
\be
\begin{array}{c} y''(\s)+\tilde\om^2y=0,\\
 (\tilde\om^2-q_1)\,y(0)+2Q_1y'(0)=0,\quad
(\tilde\om^2-q_2)\,y(\pi)-2Q_2y'(\pi)=0,\rule{0mm}{4mm}

\end{array}
\label{ybvp}\ee
that is similar to the problem (\ref{wbvp}).
The eigen-functions
$$y_n(\s)=\cos(\tilde\om_n\s+\tilde\phi_n),\qquad
\tilde\phi_n={}\mbox{arctan}\,\frac{\tilde\om_n^2-q_1}{2Q_1\tilde\om_n},
\quad n>0,\quad\tilde\om_0\equiv\th$$
of the problem (\ref{ybvp}) are mutually orthogonal with respect to another
scalar product
\be
\langle f,g\rangle^*=\int\limits_0^\pi f(\s)\,g(\s)\,d\s+
\frac{f(0)\,g(0)}{2Q_1}+\frac{f(\pi)\,g(\pi)}{2Q_2}.
\label{scal2}\ee
So the system $\exp(-i\tilde\om_n\tau)$, $n\in Z$ is complete in the class
of function $C(I)$, $I=[\tau_0,\tau_0+2\pi]$ \cite{PeSh}.

Hence, an arbitrary quasirotational disturbance $u^\mu(\tau)$
may be expanded in the Fourier series similar to Eq.~{\ref{uFour}}.
Using this expansion for the disturbance (\ref{U+u}) $u_i^\mu$ of the
velocity vectors (\ref{Urot}) we obtain with the help of
Eqs.~(\ref{gen}), (\ref{psdet}) the following expression for an arbitrary
quasirotational motion of the string with massive ends \cite{Exc}:
\be
\begin{array}{c}\displaystyle
X^\mu(\tau,\s)=X^\mu_{rot}(\tau,\s)+i\sum\limits_{n\ne0}
\Big\{\frac{\alpha_n}{\om_n} e_3^\mu\cos(\om_n\s+\phi_n)\exp(-i\om_n\tau)\\
\qquad\qquad{}+\beta_n\big[e_0^\mu f^0_n(\s)+\acute e^\mu(\tau) f_n^\perp(\s)+
ie^\mu(\tau) f_n^\parallel(\s)\big]\exp(-i\tilde\om_n\tau)\Big\}.
\rule{0mm}{4.5mm}
\end{array}
\label{osc}\ee
Here the first term $X^\mu_{rot}$ describes the pure rotation (\ref{rot})
and
$$
\begin{array}{c}
f^0_n(\s)=(q_1-\tilde\om_n^2)(2\tilde\om_nQ_1)^{-1}
\cos\tilde\om_n\s-\sin\tilde\om_n\s,\\
f_n^\perp(\s)=\Gamma_1(\Theta_n\tilde\om_n-h_nv_1)\,C_\th C_\om
-v_1^{-1}C_\th S_\om+\Gamma_1\th\Theta_n S_\th S_\om+h_n S_\th C_\om,
\rule[3.5mm]{0mm}{1mm}\\
f_n^\parallel(\s)=\Gamma_1(\Theta_n\tilde\om_n-h_nv_1)\,S_\th S_\om
+v_1^{-1}S_\th C_\om+\Gamma_1\th\Theta_n C_\th C_\om-h_n C_\th S_\om,
\rule[3.5mm]{0mm}{1mm}\end{array}
$$
 $\displaystyle
\Theta_n=\frac{2\th}{\tilde\om_n^2-\th^2},\;\;\,
h_n=\frac12\Big[\frac\th{\tilde\om_n}\Big(\frac1{v_1}+v_1\Big)+
\frac{\tilde\om_n}\th\Big(\frac1{v_1}-v_1\Big)\Big],\,$
$C_\th(\s)=\cos\th\s$, $\,S_\th(\s)=\sin\th\s$, $\,C_\om(\s)=\cos\tilde\om_n\s$,
$\,S_\om(\s)=\sin\tilde\om_n\s$, $\,\om_{-n}=-\om_n$,
$\,\tilde\om_{-n}=-\tilde\om_n$, $\,\al_{-n}=\overline{\al_n}$,
$\,\beta_{-n}=\overline{\beta_n}$.

The Fourier series (\ref{osc}) is presented here for the $3+1$\,-\,dimensional
case with two independent components of the distubance (with Fourier modes
$\al_n$ and $\beta_n$). This series may be easily generalized for an
arbitrary dimension. In this case we substitute $\al_n$ for a vector
$\al_n^\mu$ orthogonal to $e_0$, $e_1$, $e_2$.

The quasirotational disturbances (\ref{osc}) with $\beta_n\ne0$ and $\al_n=0$
are small (if $\beta_n\ll\Omega^{-1}$) harmonic planar oscillations
or stationary waves in the rotational plane.
The shape of these stationary waves in the co-rotating frame of reference
(where the axes $x$ and $y$ are directed along $e^\mu$ and $\acute e^\mu$)
is approximately described by the function
$\beta_n\big[f_n^\perp(\s)-f_n^(\s)\cos(\theta\s+\phi_1)\big]$
if the deflection is maximal.
For each $n$ this rotating curved string oscillates at the frequency
$\Omega_n=\Omega\tilde\om_n/\theta$, it has $n$ nodes in $(0,\pi)$
(which are not strictly fixed because $f_n^0$ and $f_n^\parallel$ are non-zero) and the
moving quarks are not in nodes. Note that Eq.~(\ref{osc}) describes
both the deflection of the endpoints
$\beta_n\acute e^\mu f_n^\perp(\s_i)\sin\om_n\tau$ and
their radial motion $\beta_ne^\mu f_n^\parallel(\s_i)\cos\om_n\tau$.

Among the oscillations (\ref{osc}) in the rotational
plane (planar modes) with $\al_n=0$, $\beta_n\ne0$ the main planar mode\
with $n=1$ is non-trivial\footnote{Unlike the $n=1$ mode
of orthogonal oscillations (\ref{zwave}).}. If only this mode is excited  the
motion is quasiperidical, the string at the frequency $\Omega_1=
\Omega\tilde\om_1/\theta$ ($1.5\Omega<\Omega_1<2\Omega$) slightly deflects
from pure uniform rotation keeping almost rectilinear shape. The length of the
string (distance between quarks) varies in accordance with this deflection
at the frequency $\Omega_1$. The endpoints draw curves close to ellipses both
in the co-rotating frame of reference (the pure uniform rotational position
of an end is in the center of this ellipse) and in the frame of reference $Oxy$.
In the latter case this ellipse (close to a circle) rotates in the main
rotational direction because the frequencies
$\Omega_1$ and $2\Omega$ are incommensurable numbers: $\Omega_1<2\Omega$.
The more this disturbance (its amplitude $\beta_1$) the more those rotating
ellipses differ from circles.

For the second oscillatory mode ($n=2$) the middle part of the rotating string
oscillates at the frequency $\Omega_2$, but deviations of the endpoints are
small in comparison with this amplitude in the middle part. This motion is shown
in details in Ref.~\cite{stabPRD}.

The frequencies $\om_n$ and $\tilde\om_n$ from Eqs.~(\ref{zfreq}) and
(\ref{pfreq}) are real numbers,
so the rotations (\ref{rot}) of the string with massive ends are stable
in the linear approximation.

Hence, one may consider the Fourier series (\ref{osc})
for an arbitrary quasirotational motion as the basis for quantization
of some class of motions (quasiro\-tational states) of the string with massive
ends in the linear vicinity of the solution (\ref{rot}).
Note that for the string baryon models ``three-string" (Y) and the linear
configuration ($q$-$q$-$q$) their rotational motions are unstable,
the analogs of Eq.~(\ref{pfreq}) for these models contain complex frequencies
\cite{Exc,Exclin}.

\bigskip
\noindent{\bf 3. Quasirotational disturbances and excited states of hadrons}
\medskip

The quasirotational states (\ref{osc}) of the string with massive
ends after quantization may be used for describing radial excitations of
hadrons.
The serious difficulties with quantization of the relativistic string with massive ends are connected with essentially nonlinear boundary conditions
(\ref{qq}) \cite{BN}. There were attempts to avoid these difficulties via
the rectilinearity ansatz \cite{Solov,Olss89} for the sting. But this
approach excludes practically all string degrees of freedom and reduces
all string motions to rotational ones and their simplest generalizations
\cite{PeSh}.

We suppose here another approach --- to consider the space $\cal{Q}$
of quasirotational states (\ref{osc}) for the string with massive
ends. This space contains the countable set of string degrees of freedom.

Possible applications of these string states in hadron spectroscopy depend on
their physical characteristics. Let us calculate the most important among them:
the energy $E$ and angular momentum $J$ of the quasirotational state (\ref{osc}). For an arbitrary classic state of the relativistic string
with massive ends they are determined by the following integrals
(Noether currents) \cite{BN}:
\bea
&\displaystyle
P^\mu=\int\limits_{\s_1}^{\s_2} p^\mu(\tau,\s)\,d\s
+p_1^\mu(\tau)+p_2^\mu(\tau),& \label{Pimp}\\
&\displaystyle
{\cal J}^{\mu\nu}={\!\!\int\limits_{\s_1}^{\s_2}\!}(X^\mu
p^\nu-X^\nu p^\mu)\,d\s+\sum_{i=1}^2
\Big[x_i^\mu(\tau)\,p_i^\nu(\tau)-x_i^\nu(\tau)\,p_i^\mu(\tau)\Big].&
\label{Mom}\eea
Here  $\,p_i^\mu(\tau)=m_iU_i^\mu(\tau)$, $\,p^\mu(\tau,\s)=\gamma(-g)^{-1/2}
\big[(\dot X,X') X^{\pr\mu}-X'{}^2\dot X^\mu\big]$.
In the orthonormal gauge (\ref{ort})
$p^\mu(\tau,\s)=\gamma\dot X^\mu(\tau,\s)$.

The square of energy $E^2$ equals the scalar square of the conserved
$R^{1,3}$-\,vector of momentum (\ref{Pimp}): $P^2=P_\mu P^\mu=E^2$.
In the center of mass reference frame $E=P^0$, the latter case takes place
for the expression (\ref{osc}).

If we substitute the Fourier series (\ref{osc}) into expression (\ref{Pimp})
we'll obtain the following equality for the 4-momentum:
\be
P^\mu=P^\mu_{rot}=e_0^\mu\sum_{i=1}^2[\gamma\Omega^{-1}\arcsin v_i+
m_i\Gamma_i].
\label{Prot}\ee

In other words, the energy of an arbitrary quasirotational state (\ref{osc})
equals the energy $E_{rot}$ (\ref{EJ}) of the pure rotational motion
(\ref{rot}), all quasirotational modes with amplitudes $\al_n$, $\beta_n$ yield
zero contributions.
This result is obtained in the linear approximation with respect to
$\al_n$, $\beta_n$ in the expressions for the endpoints' momenta
$p_i^\mu=m_iU_i^\mu$, for example
$$
U_1^\mu(\tau)\simeq U^\mu_{1(rot)}+\Omega v_1\Gamma_1^2\sum_{n\ne0}\Big\{
\frac{e_3^\mu\al_n}{\mbox{\footnotesize$\sqrt{\tilde\om_n^2+Q_1^2}$}}
e^{-i\om_n\tau}
-i\beta_n\big[e_0^\mu +v_1^{-1}\acute e^\mu(\tau) +
ih_ne^\mu(\tau)\big]\,e^{-i\tilde\om_n\tau}\Big\}.$$
Here $U^\mu_{i(rot)}$ is the expression (\ref{Urot}).
In this approximation the contribution of the ends in Eq.~(\ref{Pimp})
exactly compensates the string contribution
$\int p^\mu(\tau,\s)\,d\s$ for each oscillatory mode.

The equality (\ref{Prot}) of the energies of the quasirotational $E$ and
pure rotational motion $E_{rot}$ (compare with the similar result in
Ref.~\cite{AllenOV}) looks questionable: we always can add a disturbance with
nonzero energy $\Delta E$ to the pure rotation (\ref{rot})
$X^\mu_{rot}(\tau,\s)$. But there is no contradiction: the resulting motion
will be the quasirotational state (\ref{osc}) with respect to the rotation
(\ref{rot}) with the energy $E_{rot}+\Delta E$.

This property is similar to vanishing contributions of high oscillatory modes
in the Fourier series
\be
X^\mu(\tau,\s)=x_0^\mu+\frac{p^\mu}{\pi\gamma}\tau+
i\sum_{n\ne0}\frac{a_n^\mu}n\cos(n\s)\exp(-in\tau).
\label{Four0}\ee
for the massless open string \cite{BN}. In the latter case
energy of these oscillation may be found from the orthonormality conditions (\ref{ort}) (the Virasoro conditions).

In the case of the string with massive ends  we have no the Virasoro
conditions, because the orthonormality conditions (\ref{ort}) were previously
solved and taken into account in expressions (\ref{rot}), (\ref{psdet}),
(\ref{sysU}), (\ref{U+u}) in the linear approximation with respect to $u_i^\mu$.

The classical angular momentum of the quasirotational motion (\ref{osc})
is determined by Eq.~(\ref{Mom}) and in the case\footnote{Remind that inequality
$\al_1\ne0$ results in a trivial tilt of the rotational plane.} $\al_1=0$
it takes the form
\be
{\cal J}^{\mu\nu}=j_3^{\mu\nu}\bigg(J_{rot}-{}\!\sum_{n=1}^\infty\!
|\beta_n|^2\frac{\gamma\theta B_{1,n}}{4Q_1^2(\tilde\om_n^2-\theta^2)}
\cdot Z_n\bigg).
\label{Mrot}\ee
Here $J_{rot}$ is the energy angular momentum (\ref{EJ}) of the pure
rotational motion (\ref{rot}) and the following notations were used:
$$\begin{array}{c}
j_3^{\mu\nu}=e_1^\mu e_2^\nu-e_1^\nu e_2^\mu=e^\mu\acute e^\nu-
e^\nu\acute e^\mu,\qquad
B_{i,n}=4Q_i^2+(\tilde\om_n^2-q_i)^2\big/\tilde\om_n^2,\\
\displaystyle
Z_n=\pi(3\tilde\om_n^2+\theta^2)+
\sum_{i=1}^2\left[\frac{16Q_i^3}{B_{i,n}}-
\frac{v_i^2(\tilde\om_n^2-\theta^2)}{Q_i}\right].
\rule{0mm}{6mm}\end{array}$$

Note that the 2-nd order contribution $\Delta J$ (proportional to $|\beta_n|^2$)
in Eq.~(\ref{Mrot}) to the momentum $J_{rot}$ (\ref{EJ}) of the pure
rotational motion is always negative. This is natural, the rotational motion
(\ref{rot}) has the maximal angular momentum among the motions with given energy
\cite{BN}. And we consider the quasirotational states (\ref{osc}) with fixed
energy (\ref{Prot}) $E=E_{rot}$.

In the theory of massless strings the expression (\ref{Four0})
(and the similar series for the closed string) is the analog
of the planar waves expansion of the field $X^\mu(\tau,\s)$ \cite{GSW}.
So for the string with massive ends we can interpret the amplitudes
$\al_n$, $\beta_n$ in the series (\ref{osc}) as values
proportional to birth and death operators. Here the pure rotational
motion (\ref{rot}) $X^\mu=X^\mu_{rot}$ is the vacuum state $|0\rangle$.

Determine the Poisson brackets
$$\{F,G\}=\frac1\gamma\bigg[\int\limits_{0}^{\pi}
\bigg(\frac{\de F}{\de\dot X^\mu}\frac{\de G}{\de X_\mu}-
\frac{\de F}{\de X^\mu}\frac{\de G}{\de\dot X_\mu}\bigg)d\s+
\sum_{i=1}^2Q_i\bigg(\frac{\partial F}{\partial\dot x_i^\mu}
\frac{\partial G}{\partial x_{i\mu}}-
\frac{\de F}{\partial x_i^\mu}\frac{\partial G}{\partial\dot x_{i\mu}}\bigg)
\bigg],$$
[where $x_i^\mu(\tau)=X^\mu(\tau,\s_i)$] and calculate the amplitudes
$\al_n$, $\beta_n$ with the help of the series (\ref{osc}) and scalar products
(\ref{scal}), (\ref{scal2}):
$$
\al_n=-\frac{e^{i\om_n\tau}}{\Pi_n}e_{3\mu}
\langle\dot X^\mu-i\om_nX^\mu,w_n\rangle,\,\quad
\beta_n=-\frac{2Q_1e^{i\tilde\om_n\tau}}{\tilde\Pi_n\tilde\om_nB_{1,n}}
e_{0\mu}\langle\dot X^{\pr\mu}-i\tilde\om_nX^{\pr\mu},y_n\rangle^*.
$$
Here $\;\displaystyle\Pi_n=2\langle w_n, w_n\rangle=
\pi+\sum_{i=1}^2\frac{Q_i}{Q_i^2+\om_n^2},
\;\;\tilde\Pi_n=
\pi+\sum_{i=1}^2\frac{2Q_i(\tilde\om_n^2+q_i)}{B_{i,n}}$.

These Poisson brackets for the amplitudes $\al_n$, $\beta_n$:
$$
\{\al_m,\al_n\}=-i\lambda_m\de_{m+n,0},\quad
\{\beta_m,\beta_n\}=-i\tilde\lambda_m\de_{m+n,0},\quad
\{\al_m,\beta_n\}=0,
$$
where
$$
\lambda_m=\frac{\om_m}{\gamma\Pi_m},
\qquad\tilde\lambda_m=\frac{4Q_1^2(\tilde\om_m^2-\theta^2)}
{\gamma\theta B_{1,m}Z_m}.
$$
let us determine the commutators
\be
[\al_m,\al_n]=\lambda_m\de_{m+n,0},\quad
[\beta_m,\beta_n]=\tilde\lambda_m\de_{m+n,0},\quad
[\al_m,\beta_n]=0.
\label{comm}\ee

It was mentioned above that for the string with massive ends we have no
the Virasoro conditions which determine the mass spectra in the theory
of massless open (\ref{Four0}) or closed string \cite{BN,GSW}.
This results in additional degrees of freedom in the quantization procedure
and in choosing quasirotational states (\ref{osc}) from the space $\cal{Q}$
for the string with massive ends. Let us consider the simplest scheme of this
choice.

Taking into account the commutators (\ref{comm}) define the death and birth
operators in the Fock space $\cal{Q}$ in the following way:
$$
a_n=\frac1{\sqrt{\lambda_n}}\al_n,\;\;
a_n^\dagger=\frac1{\sqrt{\lambda_n}}\al_{-n},\,\;\;
b_n=\frac1{\sqrt{\tilde\lambda_n}}\beta_n,\;\;
b_n^\dagger=\frac1{\sqrt{\tilde\lambda_n}}\beta_{-n},\;\; n=1,2,\dots
$$
They obey the relations
$$
[a_m,a_n^\dagger]=[b_m,b_n^\dagger]=\de_{m,n},\quad
[a_m,b_n^\dagger]=0.$$

In these notations the momentum operator (\ref{Mrot}) takes the form
\be
{\cal J}^{\mu\nu}=j_3^{\mu\nu}\bigg(J_{rot}-\sum_{n=1}^\infty
b_n^\dagger b_n\bigg).
\label{Jbb}\ee
Here $J_{rot}$ is the $c$-number (\ref{EJ}).

The birth operators $b_n^\dagger$ acting on the vacuum state $|0\rangle$
(that is the classic rotational motion $X^\mu=X^\mu_{rot}$)
generates quasirotational states $b_n^\dagger|0\rangle$ in the space $\cal{Q}$.
The $z$-projection of the angular momentum (\ref{Jbb}) for these oscillations
is equal to $J_{rot}-1$ (here $\hbar=1$).
So we can interpret these states as hadron states with the same mass (energy)
$M\equiv E=E_{rot}$ as rotational or orbital excitations (\ref{rot}) but with
the angular momentum $J_{rot}-1$.

\bigskip
\noindent{\bf Conclusion}
\medskip

Small disturbances of rotating relativistic string with massive ends
(the quasirotational states) obtained in this paper in the form (\ref{osc})
may be used in hadron spectroscopy. String oscillations in the rotational
plane are the most interesting. On the classical level each summand with the
factor $\beta_n$ in the series (\ref{osc}) is the stationary waves with $n$
nodes.

In the framework of the approach developed on Sect.~3 these states have
the energy $E_{rot}$ and the angular momentum $J_{rot}-1$
or $J_{rot}-k$.
So they may describe radial excitations of mesons and baryons
in the frameworks of the model $q$-$qq$)
lying on the daughter Regge trajectories.
These trajectories are shown in Fig.~1 with white circles.
They have the same slope $\al'=1/(2\pi\gamma)$ as the main or parent trajectory
corresponding the orbital excitations (black circles in Fig.~1).
The latter orbital excitations are described by the rotational states
(\ref{rot}) of the relativistic string with massive ends \cite{4B,InSh}.

\begin{center}
\unitlength=1mm
\begin{picture}(122,90)
\put(10,10){\vector(1,0){100}} \put(10,10){\vector(0,1){75}}
\put(5,82){$J$} \put(110,5){$M^2$}
\put(5,20){1}\put(5,30){2}\put(5,40){3}\put(5,50){4}\put(5,60){5}\put(5,70){6}
\put(9,20){\line(1,0){2}}\put(9,30){\line(1,0){2}}\put(9,40){\line(1,0){2}}
\put(9,50){\line(1,0){2}}\put(9,60){\line(1,0){2}}\put(9,70){\line(1,0){2}}
\put(15,10){\circle*{2}}\put(30,10){\circle{2}}\put(45,10){\circle{2}}
\put(30,20){\circle*{2}}\put(45,20){\circle{2}}\put(60,20){\circle{2}}\put(75,20){\circle{2}}
\put(45,30){\circle*{2}}\put(60,30){\circle{2}}\put(75,30){\circle{2}}\put(90,30){\circle{2}}
\put(60,40){\circle*{2}}\put(75,40){\circle{2}}\put(90,40){\circle{2}}\put(105,40){\circle{2}}
\put(75,50){\circle*{2}}\put(90,50){\circle{2}}\put(105,50){\circle{2}}
\put(90,60){\circle*{2}}\put(105,60){\circle{2}}
\put(105,70){\circle*{2}}
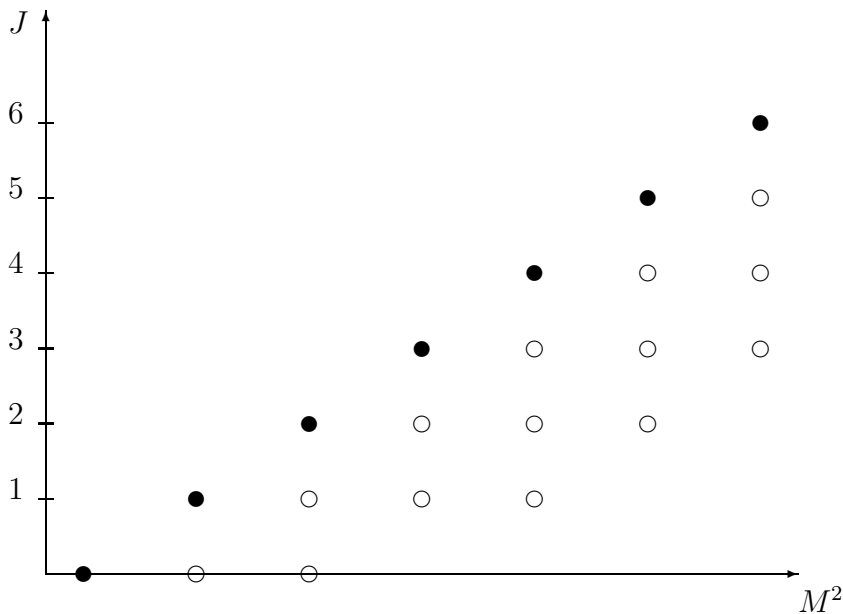
\put(6,0){Fig.\,1: Regge trajectories for rotational and quasirotational states.}
\end{picture}
\end{center}
\bigskip

This approach was developed for the string with massive ends.
We are to emphasize that it is not applicable for more complicated
string baryon models $q$-$q$-$q$ and Y (``three-string") because
the rotational motions for them are unstable and we can't quantize states
in their linear vicinity \cite{stabPRD,Exc,Exclin}.

\end{document}